\documentstyle[a4,12pt,epsf]{article}

\begin{document}

\newcommand{\Ssym}{\overline{S}_{ec}}
\newcommand{\bn}{{\bf n}}
\newcommand{\nhat}{\hat{\bf n}}
\newcommand{\dt}{\delta\tau}
\newcommand{\Gn}{G_\bn}
\newcommand{\Lff}{{\cal L}_{free}}
\newcommand{\Leff}{{\cal L}_{eff}}
\newcommand{\bX}{{\bf X}}
\newcommand{\bx}{{\bf X}}
\newcommand{\bt}{{\bf t}}
\newcommand{\G}[2]{G_{#1#2}}
\newcommand{\expect}[1]{\langle {#1} \rangle}
\newcommand{\Gp}{\G 1 1}
\newcommand{\half}{{1\over 2}}
\newcommand{\ij}{<ij>}
\newcommand{\vp}{{\vec p}}
\newcommand{\vxi}{{\vec \xi}}
\newcommand{\beq}{\begin{equation}}
\newcommand{\eeq}{\end{equation}}
\newcommand{\beqa}{\begin{eqnarray}}
\newcommand{\eeqa}{\end{eqnarray}}
\newcommand{\rf}[1]{(\ref{#1})}
\newcommand{\tdqg}{two-dimensional quantum gravity }
\newcommand{\qg}{quantum gravity }

\newcommand{\figone}{fig.2~}
\newcommand{\figtwo}{fig.1~}
\newcommand{\figtwoA}{fig.3~}
\newcommand{\figthree}{fig.4~}
\newcommand{\figfour}{fig.5~}
\newcommand{\figfive}{fig.6~}
\newcommand{\figsix}{fig.7~}
\newcommand{\figseven}{fig.8~}
\newcommand{\figeight}{fig.9~}
\newcommand{\fignine}{fig.10~}
\newcommand{\figten}{fig.1~}
\newcommand{\figeleven}{fig.1~}
\newcommand{\figtwelve}{fig.1~}

\begin{titlepage}

\begin{flushright}
Oxford Preprint: OUTP 95/07P\\
9th March 1995
\end{flushright}

\vspace*{5mm}

\begin{center}
{\Huge The Critical Exponents of Crystalline Random Surfaces}\\[15mm]
{\large\it UKQCD Collaboration}\\[3mm]

{\bf J.F.~Wheater\footnote{e-mail: jfw@thphys.ox.ac.uk}}\\
Theoretical Physics, 1 Keble Road, University of Oxford, Oxford OX1~3NP, UK

\end{center}
\vspace{5mm}
\begin{abstract}
We report on a high statistics  numerical study of the crystalline random
surface
model with extrinsic curvature on lattices of up to $64^2$ points. The critical
exponents
at the crumpling transition are determined by a number of methods all of which
are shown
to agree within estimated errors.  The correlation length exponent is found to
be
$\nu=0.71(5)$ from the tangent-tangent correlation function whereas we find
$\nu=0.73(6)$ by assuming finite size scaling of the specific heat peak and
hyperscaling.
These results imply a specific heat exponent $\alpha=0.58(10)$; this is a good
fit to
the specific heat on a $64^2$ lattice with a $\chi^2$ per degree of freedom of
1.7
  although the best direct
fit to the specific heat data yields a much lower value of $\alpha$.
Our measurements of the normal-normal  correlation functions suggest that the
model in the crumpled phase is described by an effective field theory which
deviates from
 a free field theory only by super-renormalizable interactions.

\end{abstract}

\end{titlepage}

\section{Introduction}

The crystalline random surface model has attracted considerable interest over
the past
few years.    Although such models have obvious applications in condensed
matter
physics \cite{helfrich} their role in the physics of two-dimensional
 quantum gravity is perhaps the
most intriguing.  For some time the nature of  \tdqg  coupled to conformal
matter
with central charge $c\le 1$ has been reasonably well understood
\cite{kpz,ddk};
however this understanding fails when $c>1$.  Roughly speaking when a small
number
of matter species are present  the physics is a small perturbation from that
when there
is no matter; when the number of matter species exceeds a certain value the
self coupling
of gravity through intermediary  matter fields  becomes strong and the method
of
calculation fails. (Whether the physics changes abruptly is not yet properly
understood.)

It is quite  remarkable that there is an (apparently) exact lattice formulation
of \tdqg
problems.  In this formulation the functional integral over the metric is
replaced by
summing over all triangulations of the lattice \cite{ambjorn0,Fdavid,Kazakov}
  and matter fields are provided by
placing a spin system on the lattice. At its critical point the model is the
same as
conformal matter (with $c$ equal to that of the spin system at its fixed
lattice critical
point) interacting with quantum gravity; this is known by direct calculation
 in the case of the Ising model
for example \cite{kostov}.  In this approach the simplest way to obtain
integer
$c=D$ is to  put $D$ free (that is, not self-interacting  -- they do of course
interact through
their coupling to the dynamical triangulation), massless bosonic  fields on the
lattice.
 The model is then
mathematically equivalent to one of a random surface, which is allowed to
self-intersect,
embedded in $D$  dimensions.
Unfortunately, it turns out that   such models do not lead  to a smooth
continuum limit
because there are no correlations between normals of the surface and the
continuum string
tension is always infinite \cite{durhus}.  To overcome this problem it was
proposed to
add an extrinsic curvature term to the action \cite{polyakov};   this
introduces a correlation
between normals and so might lead to a smoother surface with a finite string
tension in
the continuum limit.  The resulting model on a fixed triangulation, usually
called the crystalline
surface model, has been shown to have
a second order phase transition, as have similar models
 \cite{kantor,ambjorn,jfwrgh,ukqcd,espriu}.  Of course, since this is no longer
a free field
theory, it is clear neither that the effective field theory governing the
critical point
is a conformal field theory nor, if it is, whether it actually has $c>1$.   If
this critical point
is conformal with $c>1$ then the corresponding dynamically triangulated model
might lead
to a consistent theory of \qg interacting with  $c>1$ conformal matter.

There are several recent papers which review the status of random surface
models
{\it per se} \cite{espriu,jfwrev}  and various approaches to \tdqg
\cite{leshouches,latt94}.
The purpose of this paper is to report on a high statistics numerical study of
one
model in $D=3$ dimensions in the region of its critical point; preliminary
results which
did not use the full data set analyzed here were presented in \cite{ukqcd}.
This paper is
organized as follows.  In sections 2 and 3 we describe the model and our
simulation
methods, sections 4, 5 and 6 deal with the results and in section 7 we discuss
our
conclusions.

\section{The model}

The  crystalline surface model  approximates a continuum surface by a
(regular) triangulation with $N$ vertices labelled $i=1,\ldots N$; each vertex
has a location on
the lattice denoted by its  two-component intrinsic coordinates $\vxi(i)$ and a
location
in the $D$-dimensional euclidean embedding space $\bX(i)$.  We will sometimes
refer to the components of $\bX$ as $X_\mu$ where $\mu=1,\ldots D$. In
numerical work
it is much easier to use closed surfaces and our choice of lattice boundary
conditions
(periodic in $y$ and helical in $x$ \cite{jfwrgh}) make the surface a torus.

There are several ways of assigning an action to this system
\cite{kantor,ambjorn,baig} and for some time it appeared that they did not all
have
the same critical properties. However,  recent work \cite{espriu} has
demonstrated
that the analysis of simulations is complicated by very long autocorrelation
times
and that in fact all the models have similar critical behaviour.
  In this paper we work with the model whose action is given by
\beq   S = \beta^2 S_G  + \kappa  S_{ec}.\eeq
where $\beta$ and $\kappa$ are coupling constants.
$S_G$ is gaussian and given by
\beq S_G=\half \sum_{<ij>} (\bX(i)-\bX(j))^2 \label{SG}\eeq
where  $\ij$ denotes the link joining neighbouring points labelled by
$i$ and $j$.  $S_{ec}$ denotes  the extrinsic curvature action and is
 given by
\beq S_{ec} =-\sum_{<ij>} \nhat_{\triangle}\cdot\nhat_{\triangle'} \label{SEC}
\eeq
where $\triangle, \triangle'$ are the
 triangles sharing the link $<ij>$, $\bn_\triangle$ is the normal vector of
$\triangle$,
and   $\nhat_\triangle$  the correspond  unit vector \cite{kantor}.
The partition function  is given by
\beq Z = \prod_{i = 1}^{N}\int d^D\bx(i)\, \delta^D \biggl(\sum_{i = 1}^N
\bx(i) \biggr)\, e^{-S}, \label{ZEC} \eeq
where the delta function has been included
to remove the translational zero mode.
Expectation values are given by
 \beq \expect{{\cal O}} ={1\over Z}
 \prod_{i = 1}^{N}\int d^D \bx(i)\, \delta^D \biggl(
\sum_{i = 1}^N\bx(i) \biggr)\, {\cal O}(\{ \bx \})\,e^{-S}.  \eeq
By exploiting the invariance of $S_{ec}$ under the global rescaling
$\bX\to\lambda\bX$ it is easy to derive the results
\beqa \expect{S_G}&=& {(N-1)D\over 2\beta^2}\label{moment1}\\
	\expect{S_G^2}-\expect{S_G}^2\equiv
\expect{S_G^2}_c&=& {(N-1)D\over 2\beta^4}  \label{moment2}\eeqa
and so on; note that $\beta$ can be scaled away entirely but we will
retain it for later convenience.

At $\kappa=0$ this model is exactly soluble and describes a surface whose
mean square radius
\beq \expect{\bX^2}=\frac{1}{N}\sum_{i=1}^N\bX(i)\cdot\bX(i)\label{msqrad}\eeq
behaves as
\beq  \expect{\bX^2}\sim\log N.\eeq
Defining a ``Hausdorff" dimension $d_H$ through
\beq  \expect{\bX^2}\sim N^{\frac{2}{d_H}}\eeq
the behaviour at  $\kappa=0$ corresponds to $d_H=\infty$ and the
surface is so highly crumpled that it is space filling in arbitrary dimension
$D$;
this behaviour persists at small $\kappa$ and this phase is usually called the
``crumpled phase".  The  extrinsic curvature term,  $S_{ec}$,
  acts to suppress sharp bends in the surface and so as $\kappa$ increases
we expect the surface to become smoother.  It is now well-established
\cite{kantor,ambjorn,jfwrgh,ukqcd} that there is a second order phase
transition
at some critical value,  $\kappa_c$, of the coupling.  There are reasons for
believing that at $\kappa=\kappa_c$ the value of $d_H$ drops to 4 or less
\cite{ambjorn,jfwrgh};  as $\kappa\to\infty$ the surface  becomes smoother
and we expect $d_H\to 2$.  In fact the numerical evidence \cite{jfwrgh}
indicates
that $d_H$ goes straight to 2  at the critical point.

Although considerable effort has gone into determining the nature of the
critical point and
measuring the critical exponents
 there are
still some apparent discrepancies.   The purpose of this work was to gather
very
high statistics on lattices of moderate size in order to see if the
discrepancies
could be resolved.  As a byproduct, we have also been able to measure some
quantities
for the first time.

\section{The simulation}

There are some advantages to simulating these models using the Langevin
method with Fourier acceleration. This is largely because the action depends
only upon derivatives of the $\bX$s and not on the fields themselves; it has
been observed \cite{jfwrgh} that a Fourier accelerated simulation gives much
shorter autocorrelation times.  Although this is partly offset by the extra
computer time needed for the Fourier transforms there is still a substantial
gain.
However the method is not perfect and  critical slowing down  is still observed
near the phase transition.  The method of Fourier acceleration used in this
work is identical to that described in \cite{jfwrgh} but another aspect of the
algorithm has been changed.

In a Langevin simulation, new configurations are generated
by discretizing the Langevin time $\tau$ in steps of $\dt$; at an update
the change
in $\bX$ is given by
\beq  \delta X_\mu(i) = -\dt {\partial S\over \partial X_\mu(i)}
 +\eta(i,\tau)\sqrt{\dt}\eeq
where $\eta$ is a gaussian distributed random number satisfying
$\expect{\eta(i,\tau)\eta(j,\tau')}=2\delta_{ij}\delta_{\tau\tau'}$.
The derivative of $S_{ec}$  depends upon
\beq {\partial \nhat_{\triangle}\over \partial X_\mu(i)}=
{1\over\vert \bn_\triangle\vert} \left({\partial \bn_{\triangle}\over
 \partial X_\mu(i)}-\nhat_{\triangle}\thinspace \left(
\nhat_{\triangle}\cdot{\partial \bn_{\triangle}\over
 \partial X_\mu(i)}\right)\right) \label{nderiv}\eeq
which is non-zero if  $i$ is one of the vertices  of $\triangle$ and zero
otherwise.
If $\triangle$ has a very small area  $\vert \bn_\triangle\vert$
then it is possible that a very large update for $ X_\mu(i)$ will
be generated; there is no reason why the expression in
 brackets in \rf{nderiv}  should be particularly small when
 $\vert \bn_\triangle\vert$ is small.  Thus configurations of
the surface which have one or more triangles of very small area
are prone to generate large spikes at the next update. This
problem exists for any finite $\dt$ and any non-zero $\kappa$
and is an artefact of discretizing the Langevin equation.  To
deal with the difficulty it is necessary to modify the simulation
procedure so as to suppress the contribution
of triangles of very small area. This may be done by introducing
a small area cut-off; the resulting procedure is rather like a
hybrid algorithm. However this is computationally very inefficient
to implement and we have developed a different method.

We work with the modified extrinsic curvature action
\beq \overline{S}_{ec} =  -\sum_{<ij>} \nhat_\triangle\cdot\nhat_{\triangle'}
\exp\left(-\sigma\dt\left({1\over \vert \bn_\triangle\vert}+
{1\over \vert \bn_\triangle'\vert}\right)\right) \label{Ssym}\eeq
where $\sigma$ is some constant so that the total action for
the simulation is given by
\beq   S_{sym} = \beta^2 S_G  + \kappa \Ssym.\eeq
When $\dt\to 0$ this reduces (at least
naively) to the original action.  On the other hand at finite $\dt$
the contribution of any triangle of area much less than $\sigma\dt$
to the Langevin update is heavily suppressed.  The exponential
suppression wins over any power law from \rf{nderiv} and provides
a cut-off which is almost a theta function but has the virtue of being
differentiable; thus $\Ssym$ can be simulated by a completely standard
Langevin method. (Note that this procedure
does {\it not} suppress the creation of small triangles, it merely prevents
them generating spikes.)   Working with $\Ssym$ is also very efficient
computationally; compared to the original action $S_{ec}$ very little extra
calculation is needed.  Simple  analysis shows that
$\sigma$ is dimensionless;  in principal we could choose it to take any
value but its presence can be exploited further.  A Langevin
simulation with finite $\dt$ always renormalizes the couplings of
the bare action and in general introduces other operators as well
 \cite{lepage}.  This has the consequence that the $\bX$ configurations
that we generate do not actually yield the expectation value \rf{moment1}
correctly
because $\beta$ is renormalized; this trivial
renormalization can be computed by comparing
the measured and theoretical values of  $\expect{S_G}$.   However, even after
taking account of the renormalization of $\beta$, higher moments such as
\rf{moment2}
do not  come out as predicted  because other operators
are induced by the discretization.  We use the freedom to tune $\sigma$ to
 arrange that the measured ratio
\beq \rho={(N-1)D\over 2}\expect{S_G^2}_c/\expect{S_G}^2\label{ratio}\eeq
is close to the predicted value of 1; that is to say we can remove at least
some
of the effects induced by the discretization of the Langevin time.
   We found that it is easy to tune
$\sigma$ so that the  residual discrepancy in $\rho$
is of order 1\%.
The value of  $\sigma$ needed is approximately
independent of $\dt$, as it should be since $[\sigma]=0$, at least over the
range
$\dt=0.001$ to 0.004 which brackets the value of $\dt=0.002$  used
in our simulations as is shown in \figtwo\llap.   That the
procedure is consistent is  shown in \figone\llap;
 the value of  $\sigma$ deduced by tuning on a $16^2$
lattice works just as well on a $64^2$ lattice.
Higher moments of $S_G$ are still
affected by discretization errors; for example the connected third moment
comes out about 50\% too large (although it should be realized that it is very
difficult to measure this double-subtracted quantity accurately).

Despite the improvements due to the Fourier acceleration it is still necessary
to acquire very high statistics for reliable measurements.
For example, in the case of the specific heat  (defined in section  5)
  the total data set consisting of $T_0$ sweeps is
split into $K$ bins;  we then compute the specific heats, $C_0$
from the total data set,  and $C_k$ $(k=1,\ldots K)$  from each bin separately.
We
define  $K_{bin}$ to be the largest value of $K$ for which
\beq \left\vert1- {1\over C_0 K}\sum_{k=1}^KC_k \right\vert <0.01\eeq
and then
\beq T_{bin}=T_0 K_{bin}^{-1}\label{Tauto}.\eeq
Subsequent analysis is then done using bins of size $T_{bin}$;  in this way
we ensure that the statistical errors in the measured specific heat are larger
than the
thermalization errors.
The values of $T_{bin}$ found, measured
in numbers of second order Runge-Kutta iterations of our algorithm, are shown
 in \figtwoA\llap; as can be seen  $T_{bin}$ appears to scale roughly like
$\sqrt N$
which is consistent with the behaviour found in \cite{jfwrgh}.
We have run our simulations for at least  $10T_{bin}$ iterations except in the
case of
crucial $\kappa$ values where we have run for much longer; in every case the
first
$2T_{bin}$ iterations  are discarded.  Thus we have
accumulated about
$10^8$ iterations of the algorithm in total.

\section{Tangent-Tangent Correlation Functions}

 Much of the
information about  the long distance properties of the system is
contained in the tangent-tangent correlation functions.
Let
\beqa \bt_1(\vxi)\equiv\bt_1(\xi_1,\xi_2) &=& \bX(\xi_1+1,\xi_2) -
\bX(\xi_1,\xi_2)\\
\bt_2(\vxi)\equiv\bt_2(\xi_1,\xi_2) &=& \bX(\xi_1+1,\xi_2+1) -
\bX(\xi_1,\xi_2)\eeqa
then the correlation functions $\G i j$ are defined by
\beq {\G i j}(\xi)= {1\over N}\sum_{\xi_2}
 \expect{ \bt_j(\xi,\xi_2)\cdot\bt_i(0,\xi_2)}\eeq
On a periodic lattice it follows that \cite{jfwrgh}
\beq  \sum_{\xi} \Gp(\xi)= {1\over N}\sum_{\xi_2}
 \expect{ \left(\sum_{\xi}\bt_1(\xi,\xi_2)\right)\cdot\bt_1(0,\xi_2)} =0\eeq
and therefore that $\Gp(\xi)$, which is clearly positive for small
enough $\xi$, must become negative at large $\xi$.  Thus there must be some
intermediate  value, $\xi_0$, for which  $\Gp(\xi_0)=0$.
 As shown in \cite{ambjorn,jfwrgh}
the finite size scaling behaviour of  $\xi_0$ with $L=\sqrt N$
is related to the Hausdorf dimension of the surface:
\begin{enumerate}
\item $\xi_0 \to constant $  as $L\to \infty$;  this is the crumpled phase
 and has $d_H=\infty$.
If there is only one mass scale then $\xi_0\propto m^{-1}$
 where $m$ is the mass gap.
\item   $\xi_0 \to L^{1-\eta}$, $0\le \eta\le 1$,  as $L\to \infty$; then
 $d_H=4(2-\eta)^{-1} \le 4$ and the system is not crumpled.
\end{enumerate}
In this work we are examining the crumpled phase so 1)  is the relevant
scenario.  When $L$ is finite, $\xi_0$  will saturate as $\kappa$ is increased
towards the critical point because it cannot in any case exceed $L/4$ and so we
expect to see finite size effects.  The values of $\xi_0$ extracted from
$\Gp$ on $16^2$, $32^2$ and $64^2$ lattices are plotted in
\figthree  and show these effects clearly.
In order to extract the true value of $m$ near the critical point it is
necessary
to have a model for the finite size effects. Previous work \cite{jfwrgh,ukqcd}
has shown that the data for $\Gp$ at long distances in the crumpled phase
is quite well accounted for by  supposing that it is described by the effective
Lagrangian
\beq {\cal L}_{free} = {1\over A(\kappa)}\left( -{1\over
2}m(\kappa)^2\bX\cdot\nabla^2\bX +
\bX\cdot\left(\nabla^2\right)^2\bX\right) \label{Leff} \eeq
which leads to the momentum space correlation function
\beq G_{ij}(\vp)=\expect{\bt_i(\vp)\cdot\bt_j(-\vp)}= A(\kappa)
 {p_ip_j\over p^2 \left(p^2+m(\kappa)^2\right)} \label{XXff}.\eeq
 Using this asumption
we can remove the finite size effects  by fitting  \rf{XXff} to the data in the
region where $\Gp(\xi) < 0$.  But, how good is the assumption?

In \figfour  we show a sequence of measured $\Gp$ at different $\kappa$ values
on a $64^2$ lattice together with the fit described above.
  As can be seen, at large distances the data is well fitted by
\rf{XXff}  right up to the phase transition.  The last picture at $\kappa
=0.82$
shows completely different behaviour; the surface is no longer in the crumpled
phase.  This phenomenon provides us with an upper bound on $\kappa_c$ which can
be used  to constrain the fits for critical exponents.
    No error bars are shown on the data. The errors are smaller than the
symbols
on the figure but  the correlation
function at successive $\xi$ values is highly correlated;  the fits of
\rf{XXff} were made
by minimizing the $\chi^2$ function obtained using a correlated error analysis
and SVD.
However we do not actually know the true probability distribution of this data
and so
the errors on the values of $A$ and $m$ were obtained using a bootstrap
procedure;
all the values so obtained are listed in Table 1 together with the actual
number of
degrees of freedom (as determined by the SVD) contributing to the fits.
In fact the data is of high quality and, as can be seen from Table 1, the
number of
contributing degrees of freedom is large so our procedure should be reliable.
There is however one possible source of systematic error. At short enough
distances
the two point function is not well described by \rf{XXff} and so it is
necessary to
decide the minimum $\xi$ at which to attempt the fit.  Obviously there are
different
ways of doing this and it is not clear that they will give the same results. To
check on this
we tried some other procedures than fitting to the region where  $\Gp(\xi) <
0$.  The
$\tilde m$ column in Table 1 shows the results of fitting in the range $\xi\ge
10$ for
all $\kappa$ values.  As can be seen the $\tilde m$ values are always very
close to the $m$
values but the estimated statistical errors are somewhat larger (probably
because the
number of contributing degrees of freedom is generally smaller).

\begin{table}\caption{${\cal L}_{eff}$ parameters deduced from correlation
functions.}
\begin{tabular}{|c|c|c|c|c|c|}\hline
$\kappa$&$m$&$A$&d.o.f.&$\tilde m$&$g$\\\hline\hline
 0.7400  &  .348(2) & 8.95(7) & 15 & -- & -- \\
 0.7500  & .313(2)  & 8.64(10) &16 & -- & -- \\
 0.7600 &   .273(2) & 8.02(8) &17 & -- & 5.8\\
 0.7700 &   .241(2) & 7.66(5) &18 & -- & 4.7 \\
 0.7750  &  .220(2) &  7.42(6) &17 & -- & --\\
 0.7775   & .203(2) & 7.21(8) &18 & .204(4)& --\\
 0.7800 &  .201(2) & 7.28(6) & 18 & .203(3)& 3.7\\
 0.7825 &  .192(3) & 7.10(8) &19 & .192(3)&--\\
 0.7850  & .174(2) & 6.84(5) & 19 & .174(2)&--\\
 0.7875 &  .167(2) & 6.76(6) &19 & .165(3)&--\\
 0.7900 &  .155(3) & 6.59(7) & 19 & .153(4)&3.0\\
 0.7925 &  .144(2)  & 6.47(6) &19 & .143(3)&--\\
 0.7950 &  .130(2) & 6.29(3) &19 & .129(2)&--\\
 0.8000 &  .110(3) & 6.06(8) &19 & .108(5)&1.8\\
 0.8050  & .078(5) & 5.70(10) &18 & .076(7)&--\\
 0.8075 &  .057(4) & 5.47(7) &19 & .055(6)&--\\
 0.8100  & .039(7) & 5.36(8) &18 & .037(12)&1.1\\
 0.8110  &  .030(4) & 5.33(6) &19 & .029(11)&--\\
 0.8125  &  .011(11) &  5.27(3) &19 & .012(10)&--\\ \hline
\end{tabular}
\end{table}

The measured values of $m$ for $64^2$ lattices are plotted in
\figfive\llap.  We have much higher statistics than in \cite{ukqcd} and,
crucially,
reliable measurements for $\kappa$ above 0.8.     The consequence of this
is that the best fit of the data to
\beq m= a(\kappa_c-\kappa)^\nu \eeq
quoted in \cite{ukqcd} ($\kappa_c=0.821(5)$, $\nu=0.89(7)$)
 is now seen to be a poor fit to the data very  close to the critical point.
Analysis of the new data set gives
\beqa \nu&=&0.71(5)\\
\kappa_c&=&0.814(2) \label{nuk}\eeqa
with a $\chi^2$ per degree of freedom of 1.75.  This fit is shown by the
middle of the three lines in the figure  while the outer lines
 show the  fits with $\nu$ taken  such that  the $\chi^2$ per degree
 of freedom is 2.75.  If the $\tilde m$ values are fitted instead  then we find
\beqa \nu&=&0.68(10)\\
\kappa_c&=&0.813(3) \label{nuk1}\eeqa
with a $\chi^2$ per degree of freedom of 0.98.
The central values are in good agreement with \rf{nuk} but the errors are
larger reflecting the larger errors in the  $\tilde m$ values.

  It is encouraging that
this new direct measurement of $\nu$ is apparently in much better  agreement
with the result obtained by finite-size scaling of the specific heat peak
\cite{ukqcd,bengt} but salutary in that  it is necessary
to work very close to the critical point  to obtain this agreement.

\section{The specific heat}

There is a  potential ambiguity in our treatment of the specific heat depending
upon whether we calculate  it with the original action $S_{ec}$ \rf{SEC}
 or the action  $\Ssym$ \rf{Ssym} used
in  the actual simulation; we have analysed both with very similar results for
critical exponents so here we will just present the results  for  $\Ssym$.
The specific heat arising from the gaussian part of the action is
analytic as a function of $\kappa$ so we subtract it  and define the specific
heat to be
\beq C=\expect{S_{sym}^2}-\expect{S_{sym}}^2-\beta^4\left(\expect{S_G^2}-
\expect{S_G}^2\right)\label{sphtdef}\eeq
%
 Our data is shown  in \figsix\llap.  The heights and locations of the
peaks for various system sizes can be analyzed using the standard
 finite size scaling relations
\cite{fisher}
\beqa  C_{max} &=& a +b L^\omega\\
\kappa_{max} &=& \kappa_c+b' L^{-{1\over \nu}} \label{fss}\eeqa
where $a$, $b$ and $b'$ are constants and it is expected that
$\omega=\alpha/\nu$.
    Unfortunately our
 systems have two different  shapes (this is forced upon us by our
FFT implememtation which only allowed $N$ to be a power of 2) and the
coefficients $b$ and $b'$ might depend upon the shape;   ignoring this for the
moment
and taking $L=\sqrt N$ we find
\beq \omega=0.74(20).\eeq
This can be compared with the value  $\omega=0.76$ deduced
 from the symmetric lattice data alone  so we conclude that at this
level of accuracy it is justifiable to include all the data in order to
obtain an error estimate.       Although the central value for $\omega$
differs  from \cite{bengt}  who found $\omega=1.11(11)$
it is clear that with present
statistics the error on $\omega$ is rather large.
The errors on the measured  values of $\kappa_{max}$  are sufficiently
large for the smaller lattice sizes
 that \rf{fss} is not very effective for deducing the correlation
length exponent; however the data is consistent with the value
$\nu=0.71(5)$  derived from the correlation functions.

Assuming the hyperscaling result $\alpha=2(1-\nu)$ and the finite
size scaling result  $\omega=\alpha/\nu$ we find that
\beqa \nu&=&0.73(6)\\
\alpha&=&0.54(10)\eeqa
Using the measured value of   $\omega$  and
our independent measurement of $\nu$ \rf{nuk} we obtain
\beq \alpha=\omega\nu=0.53(14).\eeq
Alternatively $\alpha$ can be found from \rf{nuk}
by assuming  hyperscaling
in which case
\beq \alpha=2(1-\nu)=0.58(10)\label{hss}.\eeq
The agreement between these methods of determination is
very satisfactory but the value of $\alpha$ is much larger than
that  estimated by a direct fit to the specific heat in \cite{ukqcd}.
However the shape of the specific heat curve does not determine
$\alpha$ very accurately and  a fit of the form
\beq C=a+b(\kappa_c-\kappa)^{-\alpha} +c\kappa\label{fitto64}\eeq
to the full $64^2$ data  for $\kappa\le 0.795$ with $\kappa_c=0.814$
\rf{nuk} and $\alpha=0.58$ \rf{hss} yields a $\chi^2$ per degree of freedom
of 1.7; the  line through the $64^2$ data points in \figsix shows this
fit.    Thus we can now conclude, on the basis of much greater statistics
than in \cite{ukqcd}, that all  the different methods of analyzing the critical
behaviour are broadly consistent.

\section{Normal-Normal Correlation Functions}

The normal-normal correlation function defined by
\beq   \Gn (\vxi) = \expect{ \bn(\vxi)\cdot\bn(0)} \eeq
where
\beq \bn(\vxi)=\bt_1(\vxi)\times \bt_2(\vxi)\eeq
 is interesting because it tells us something about
the four-point couplings in the theory.  If the model were described  exactly
by the free field Lagrangian \rf{Leff}  then  this correlation function  would
be
given by
\beq  \Gn(\vxi) = {2\over 3}\thinspace \left( \Gp(\vxi) \G22(\vxi)-
\G12(\vxi)\G21(\vxi) \right)
\label{NNff}\eeq
At large distances this gives
\beq \Gn(\vxi)\sim -{1\over\vert\vxi\vert^4} \label{NNas}\eeq
Corrections to this behaviour can take two generic forms.  Firstly the power of
the
scaling law at large  $\vert\vxi\vert$ may change or the coefficient of the
$\vert\vxi\vert^{-4}$ behaviour could be different from that predicted by
\rf{NNff};
this is the behaviour that would be obtained if a relevant operator other than
those in
$\Lff$ were present.  On the other hand an irrelevant operator  would generate
deviations
from  \rf{NNff}  which fall off faster than  $\vert\vxi\vert^{-4}$ at large
distances.
For example, adding interaction terms so that the effective Lagrangian is
\beq {\cal L}_{eff}= \Lff + {g(\kappa)\over A(\kappa)^2}
\vert\partial_1 \bX\times\partial_2 \bX\vert^2, \label{Lint} \eeq
where $g$ is a dimensionless constant,
gives a super-renormalizable theory and at large distances the induced
correction to
$ \Gn(\vxi)$  falls off as  $\vert\vxi\vert^{-6}$.   In fact our data is fully
consistent
with the picture that in the crumpled phase there are only  interaction terms
of
a super-renormalizable nature.

  In \figseven we show the measured  $ \Gn(\vxi)$  for $\vxi$ lying along the
x-axis
at $\kappa=0.77$ on a $64^2$ lattice; the ``free" line shows  the form
predicted by
$\Lff$ using the parameters $A,m$ determined by the  fits to $\Gp$ described in
section 4.
Clearly at distances  $\vert\vxi\vert > 18$ the predicted form is consistent
with the measured
one while at shorter distances there is a large discrepancy.  This discrepancy
can largely be
explained by assuming a four-point coupling of the form \rf{Lint}  and
computing the diagram
shown in \figeight\llap.    The  ``interacting" line in \figseven shows the
computed   $ \Gn(\vxi)$
obtained in this way  with $g$ tuned to 4.7  which gives the best fit.    As
can be seen this
fits the
data much better at short distances but makes little difference  to the long
distance tail.  As we move
closer to the critical point it is necessary to look at larger
 distances to see the asymptotic behaviour.  The behaviour of $ \Gn(\vxi)$
 at $\kappa=0.79$  is shown in \fignine\llap.
 On a $64^2$ lattice there is now a discrepancy
between  the data and the free field prediction out to distances
$\vert\vxi\vert \simeq 25$.
the picture also shows what happens on a     $128^2$ lattice.   Unfortunately
the data
is not very good but there is no evidence for a systematic deviation from
\rf{NNff}
  at large  distances.  The two curves showing the predictions of  \rf{Lint} on
 $64^2$  and  $128^2$ lattices also show that there are significant finite size
effects
on the $64^2$ lattice at this $\kappa$; clearly using the $128^2$ data to
estimate $g$
would yield a smaller value.

The values of $g$ deduced at some $\kappa$ values are shown in Table 1.  These
values should be treated with some care because it is clear that there is a
systematic
discrepancy between the data and the prediction even when the interaction term
is included; this makes it difficult to estimate an error for the values of $g$
quoted
but it should be taken to be at least of order 20\%.   It can be seen from the
table
that the value of $g$  is apparently falling as $\kappa$ approaches $\kappa_c$
so there
is no indication for the appearance of a relevant perturbation away from $\Lff$
as
the system  approaches the critical point.

Finally we note that because the corrections to $\Lff$ are apparently
super-renormalizable their presence does not affect our analysis of the  two
point function. At $O(g)$ only the $m$ parameter is renormalized and the
form of the two point function is unaffected;  $O(g^2)$ contributions will
affect $\G i j$ at short distances but it is clear that there is a substantial
range of $\xi$  for which they can safely be ignored.

\section{Discussion}

We have determined the critical exponents $\alpha$ and $\nu$ by several
different methods involving measurements of correlation functions and of the
specific heat and making various theoretical assumptions.  All the results
obtained
are in good agreement with each other.  However in order to obtain this
 agreement it is necessary to make measurements with small statistical errors
very close to the critical point and the nature of the analysis is slightly
delicate.
To obtain manifestly assumption-free measurements of the exponents  would
clearly require the use of much bigger lattices and need very much larger
amounts
of computer time than we have used.

For the first time, and as a direct consequence of our high statistics, we have
been
able to make a convincing measurement of the normal-normal correlation
function.
The consistency of our results with the idea that the effective field theory
at the
critical point is $\Lff$  means that there is still an unresolved puzzle
concerning
these models.  At the critical point the free energy density $f$ on a strip of
infinite length and finite width $L$ with periodic boundary conditions is
expected
to behave as
\beq f=f_\infty +{\pi A\over 6L^{\rho+1}}+\ldots\label{strip}\eeq
If the critical point is conformally invariant then $\rho=1$ and $A$
is
 the central charge \cite{cardy}.  Attempts have been made to measure this
effect \cite{renken,bengt}; assuming $\rho=1$ these workers all found that
$A<1$.  On the other hand $\Lff$ (which is {\it not} conformally invariant)
predicts that $\rho=1$ and $A=6$.  Whether  or not the critical point is
conformally
invariant these two results are  apparently inconsistent.  However these
measurements of $c$
are all done on strips that are rather narrow, the widest strip being  only
twenty lattice
spacings wide. From our measurements of $\G_n$ at $\kappa=0.79$ we know that
physics differs substantially from the asymptotic behaviour implied by $\Lff$
(fig.10).
Thus, if one believes the scenario implied by $\Leff$ then  at $\kappa$ values
less than
0.02 away from the critical point one would  not expect to see the correct
asymptotic
behaviour of \rf{strip}.  In fact these measurements are done  as close to
the critical point as possible  so it may be that they are misleading.
Of course, it is equally possible that the $\Leff$ scenario is incorrect; our
measurements
are consistent with it but they are not accurate enough at very large distances
to rule
out other possibilities.   To make further progress it will probably  be
necessary to
make renormalization group studies of this model.

\subsection*{}

I am grateful to Domenec Espriu for his comments on this work.
These simulations were done
on  a Meiko i860 Computing Surface  and   some of the i860 assembler routines
 were supplied by C.Michael.   Stephen Booth  of the
Edinburgh Parallel Computing Centre, Mike Brown of the
Edinburgh University Computing Service  and Peter Stephenson
ensured the smooth running of this project.
This research was supported by the UK Science and Engineering
Research Council under grants GR/G~32779, GR/G~37132, GR/H~49191, GR/H~01243,
 GR/H~00772, GR/J~21200 and GR/J~21354 by the
University of Edinburgh and by Meiko Limited.

\centerline{\bf Figure Captions}
\smallskip
\begin{enumerate}

\item{The value of $\sigma$ \rf{Ssym} required at different $\kappa$
values to yield $\rho=1$ \rf{ratio} as determined on a $16^2$ lattice.}

\item{The ratio $\rho$ \rf{ratio} determined on $32^2$ and $64^2$ lattices.}

\item{The bin size,  $T_{bin}$, estimated from the specific heat  \rf{Tauto} as
a function
of $\kappa$.}

\item{The location of the zeroes of the correlation function $\G 1 1$ as a
function of
$\kappa$ for different lattice sizes.  Errors are smaller than the size of the
symbols.}

\item{The correlation function $\G 1 1$ in the region where it is negative
 for different $\kappa$ values on
a $64^2$ lattice. The dashed lines show a fit of the form of \rf{XXff} with the
parameters
given in Table 1. Errors are discussed in the text.}

\item{The mass gap $m$ (see Table 1) plotted as a function of $\kappa$.  The
three lines
are discussed in the text.}

\item{The specific heat as a function of
$\kappa$ for different lattice sizes. The line is the  fit of \rf{fitto64} to
the
$64^2$ results with the constraint that $\kappa_c=0.814$.}

\item{The correlation function  $\Gn$ at $\kappa=0.77$ on a $64^2$ lattice.
The ``free" line is that predicted by \rf{NNff} while the ``interacting" line
is that
predicted by \rf{Lint} with the parameters given in Table 1.}

\item{Correction to $\Gn$ at $O(g)$. The open circles represent $\bn$ sources,
the filled circle a vertex of the form \rf{Lint} and the lines $\G i j$s.}

\item{The correlation function  $\Gn$ at $\kappa=0.79$ on a $64^2$ and
$128^2$  lattices.
The ``free" line is that predicted by \rf{NNff} while the ``interacting" lines
are those
predicted by \rf{Lint} with the parameters given in Table 1.}

\end{enumerate}

\newlength{\figheight}
\begin{figure} \caption{}
\setlength{\figheight}{4 truein}
\epsfysize=\figheight
\epsfbox{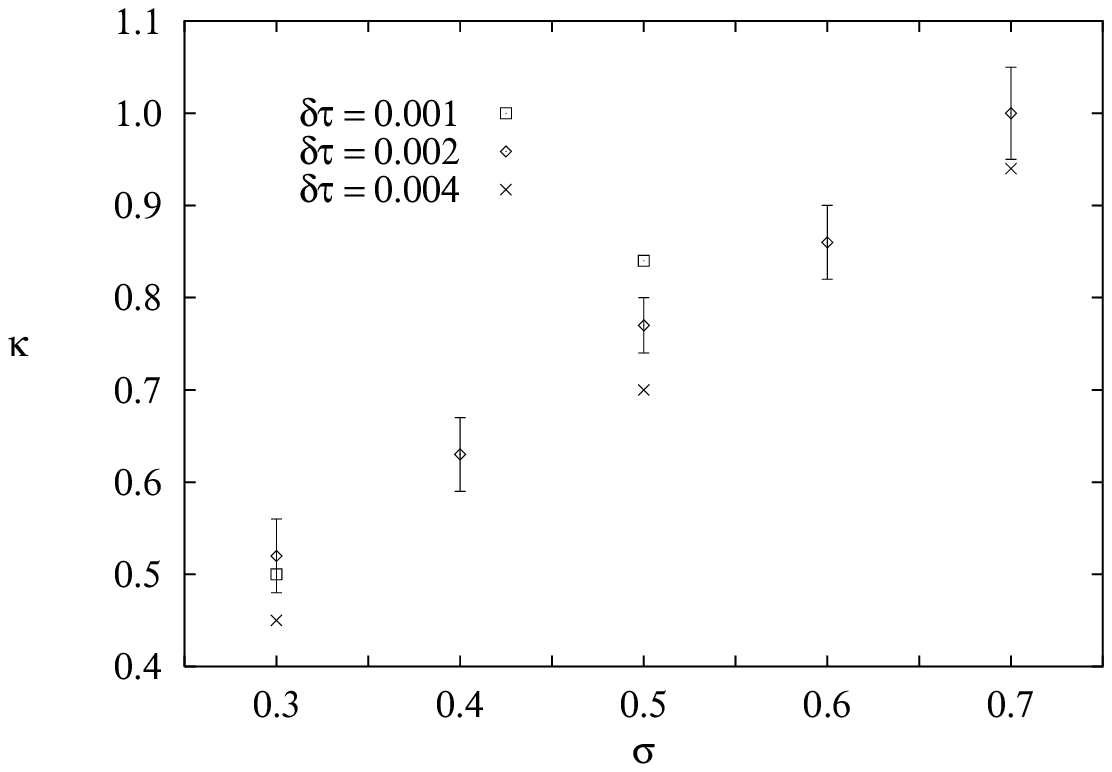}
 \end{figure}

\begin{figure}  \caption{}
\setlength{\figheight}{4 truein}
\epsfysize=\figheight
\epsfbox{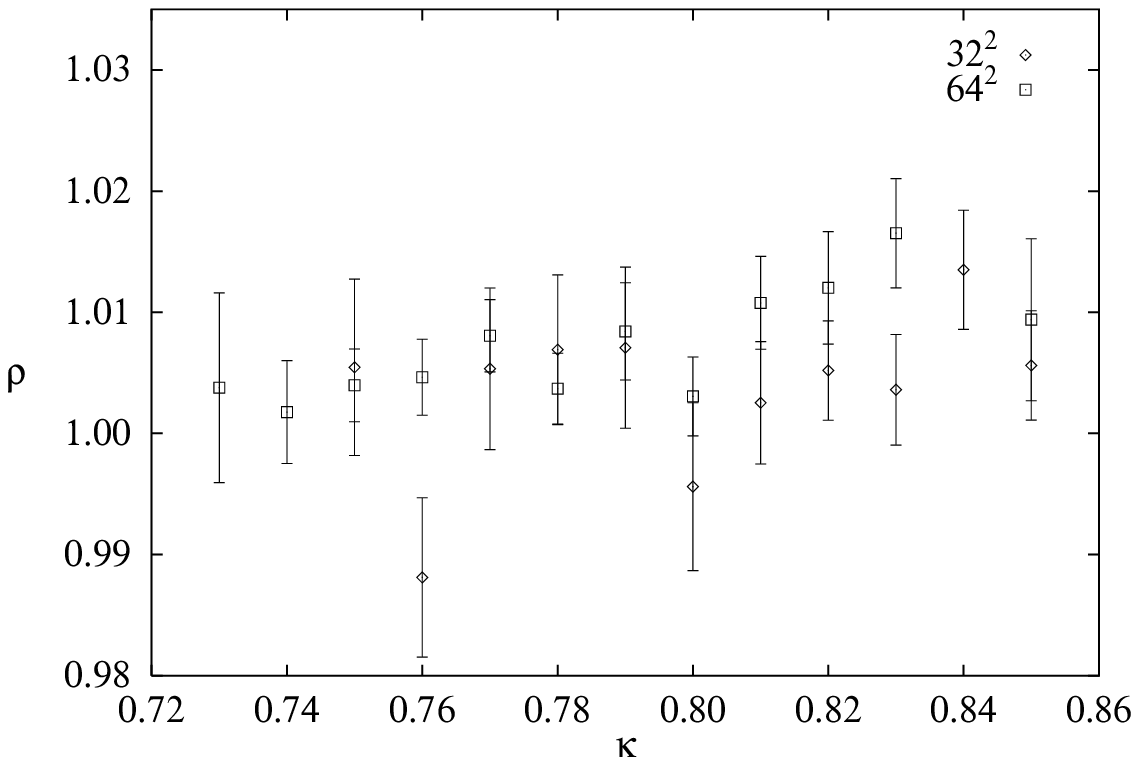}
 \end{figure}

\begin{figure}  \caption{}
\setlength{\figheight}{4 truein}
\epsfysize=\figheight
\epsfbox{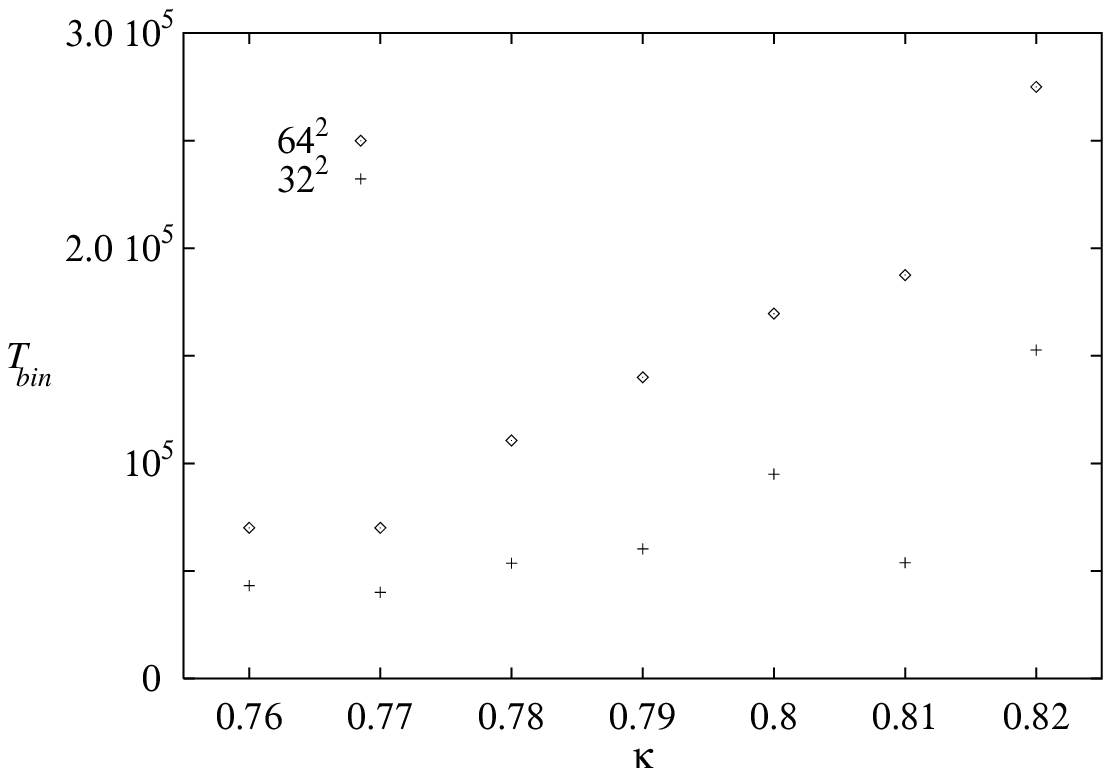}
 \end{figure}

\begin{figure}  \caption{}
\setlength{\figheight}{4 truein}
\epsfysize=\figheight
\epsfbox{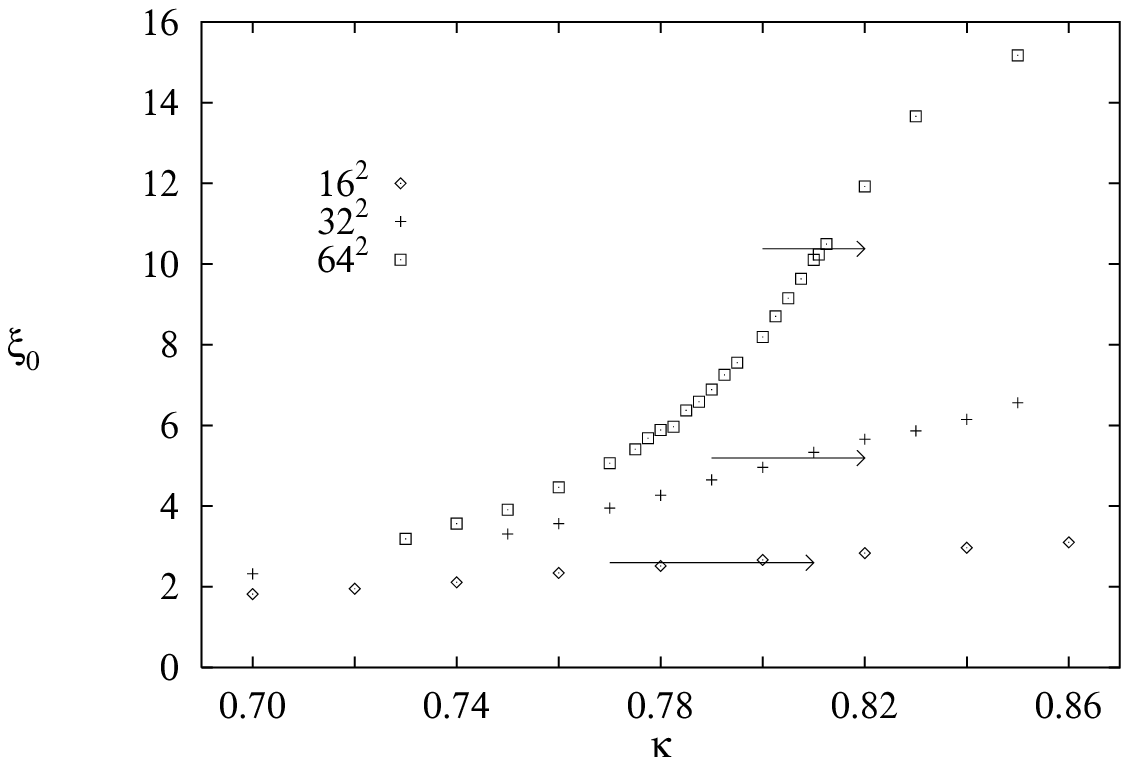}
 \end{figure}

\begin{figure}  \caption{}
\setlength{\figheight}{4 truein}
\epsfysize=\figheight
\epsfbox{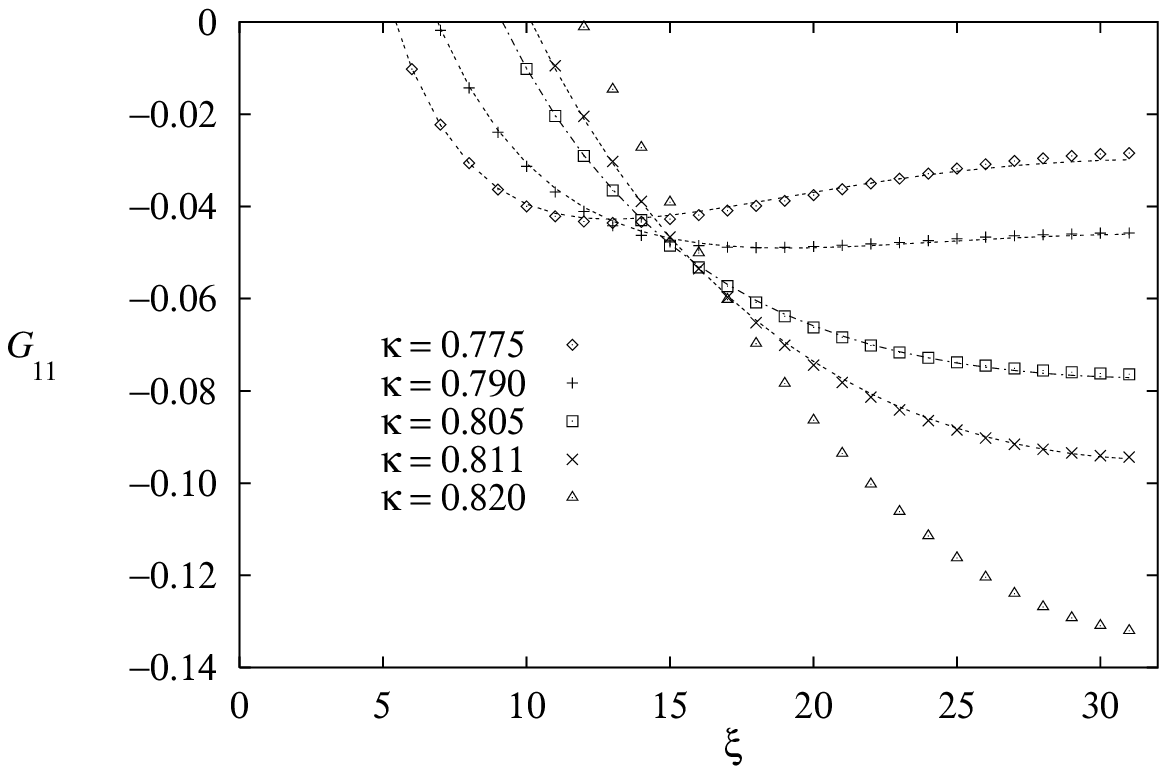}
 \end{figure}

\begin{figure} \caption{}
\setlength{\figheight}{4 truein}
\epsfysize=\figheight
\epsfbox{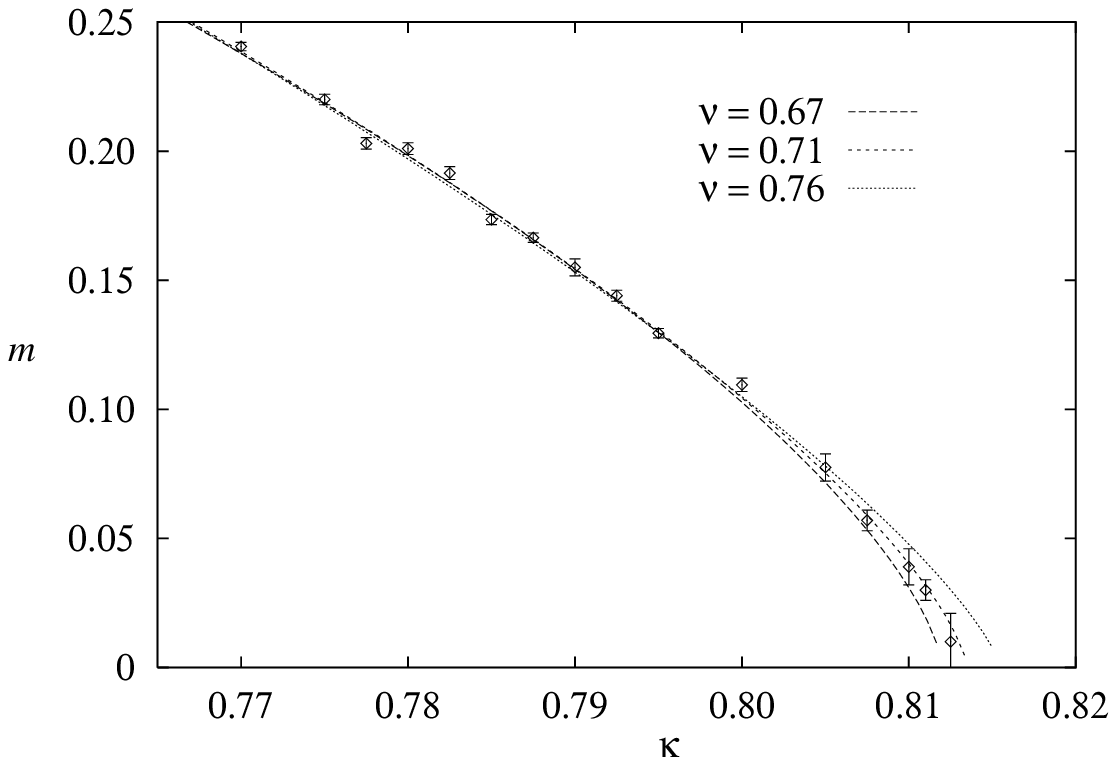}
 \end{figure}

\begin{figure} \caption{}
\setlength{\figheight}{4 truein}
\epsfysize=\figheight
\epsfbox{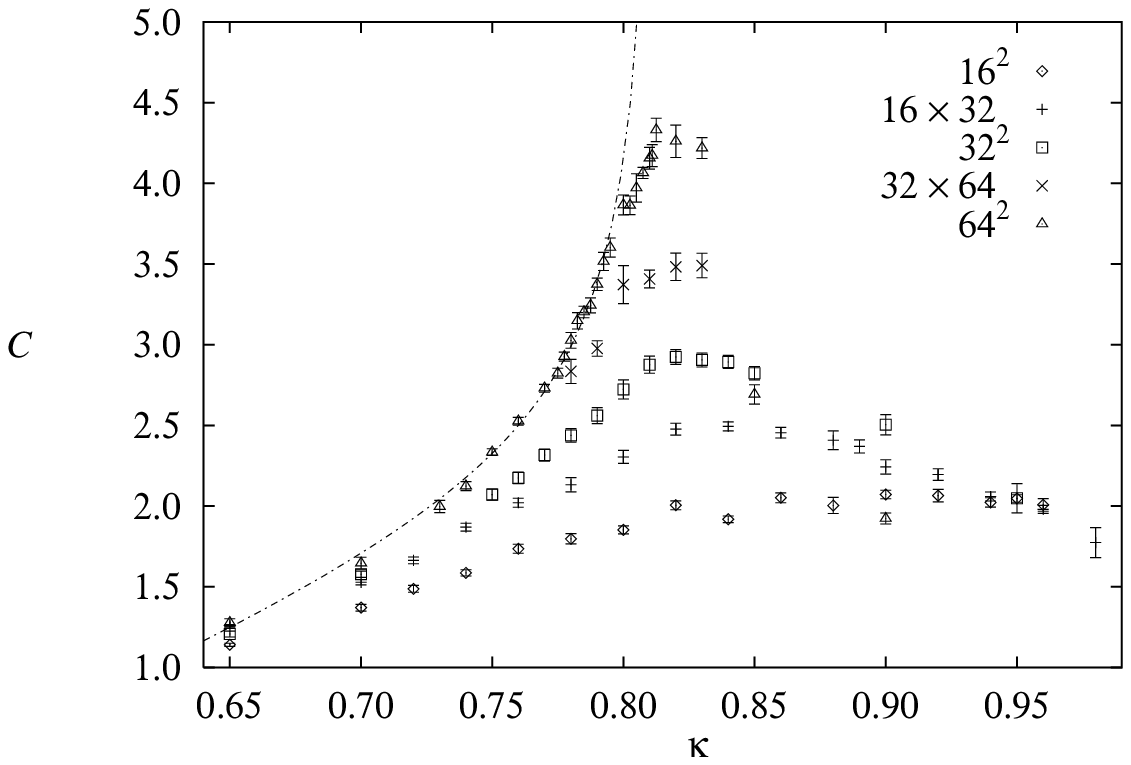}
 \end{figure}

\begin{figure} \caption{}
\setlength{\figheight}{4 truein}
\epsfysize=\figheight
\epsfbox{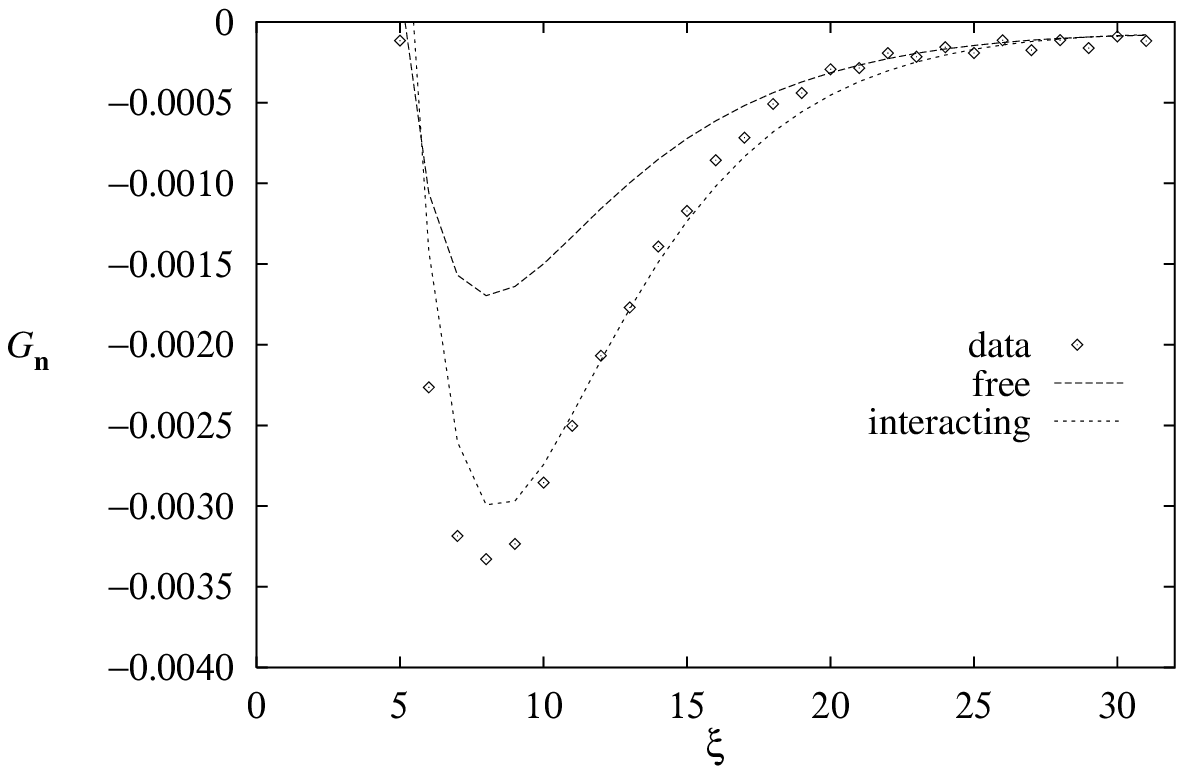}
 \end{figure}

\begin{figure}  \caption{}
\medskip
\setlength{\figheight}{0.7 truein}
\epsfysize=\figheight
\epsfbox{loopfig.eps}
 \end{figure}

\begin{figure}  \caption{}
\setlength{\figheight}{4 truein}
\epsfysize=\figheight
\epsfbox{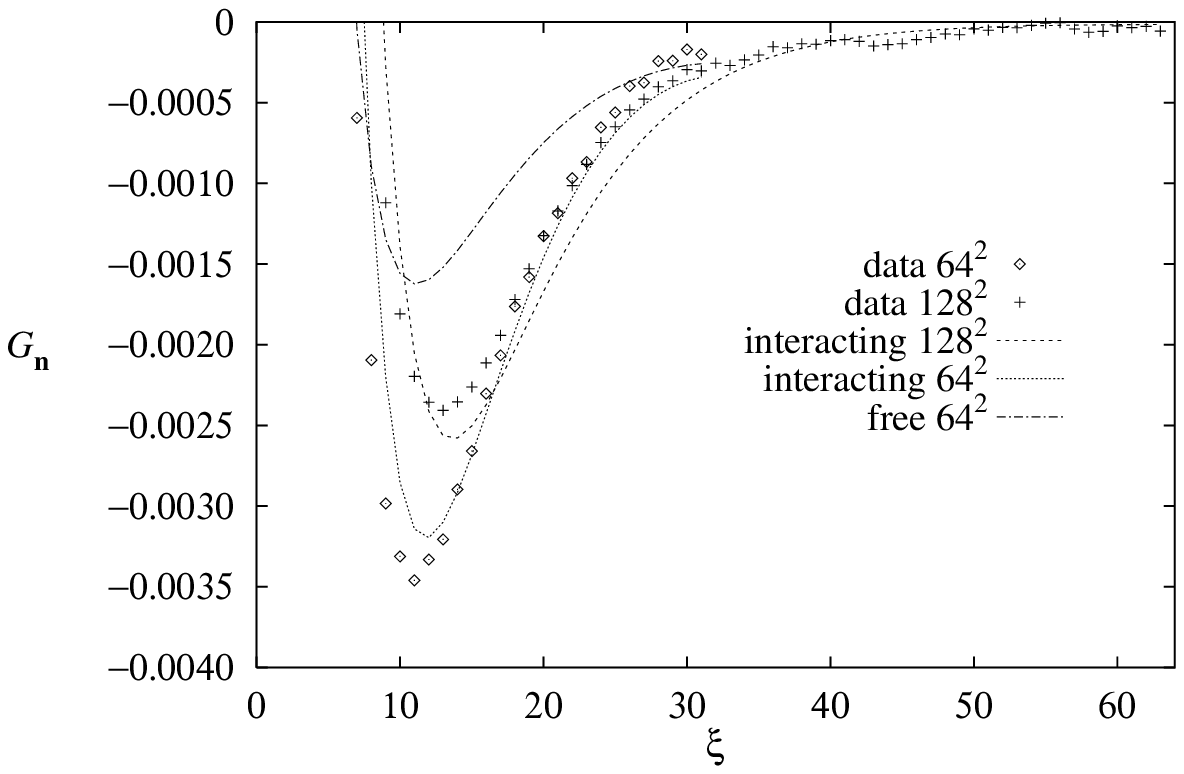}
 \end{figure}

\end{document}